\documentclass[a4paper,article,10pt]{memoir}
\usepackage[english]{babel}
\usepackage{kaeiog}
\usepackage{dcolumn}
\usepackage{bm}
\newfont{\mm}{msbm10}

\begin{document}
%
%
%
\title{%
Breakthrough in Interval Data Fitting\\
II.~From Ranges to Means and Standard Deviations} 
%
%
\author{%
    Marek W.~Gutowski
    \\ \bigskip\small
Institute of Physics, Polish Academy of Sciences, 02--668 Warszawa, Poland,
email:~{gutow@ifpan.edu.pl}
    }
%
%
    \maketitle
    \thispagestyle{empty} 
%
%
%
\begin{abstract}
~---~Interval analysis, when applied to the so called problem of experimental data
fitting, appears to be still in its infancy.\  Sometimes, partly because of the unrivaled
reliability of interval methods, we do not obtain any results at all.\  Worse yet, if this
happens, then we are left in the state of complete ignorance concerning the unknown
parameters of interest.\  This is in sharp contrast with widespread statistical methods
of data analysis.\  In this paper I show the connections between those two approaches:
how to process experimental data rigorously, using interval methods, and present
the final results either as intervals (guaranteed, rigorous results) or in a more familiar
probabilistic form:\ as a~mean value and its standard deviation.
\end{abstract}

\medskip
This article is a~companion paper to \cite{part1} and is meant to be its extension, but otherwise it
is self-contained.\ This is why we don't repeat everything here,
except for the most important thing: a~correct way to bound the distances between uncertain
experimental values and the corresponding theoretical predictions of thereof.\

\CCLsection{The goals of experimental data processing}
The problem in front of us may be stated as follows.\  We have $N$ experimental data points,
labelled as ${\mathbf m}_{1}, \ldots, {\mathbf m}_{N}$ (measurements), each one obtained in different
conditions ${\mathbf x}_{j}$,\ $j=1, \ldots, N$,\ (called {\em environments\/} from  now on), so that each
${\mathbf m}_{j}={\mathbf m}_{j}({\mathbf x}_{j})$.\  In addition, we have a theory, ${\mathcal T}$,
predicting the behavior of the investigated phenomenon in various environments.\  ${\mathcal T}$ is
characterized by $k$~\ ($k<N $)\ unknown parameters, ${\mathbf p}_{1}, \ldots\, {\mathbf p}_{k}$, so formally we
can write: ${\mathcal T}\left({\mathbf p}_{1}, \ldots, {\mathbf p}_{k}, {\mathbf x}_{j}\right) = {\mathbf t}_{j}$.\
In words: when the (yet) unknown parameters have values ${\mathbf p}_{1}, \ldots, {\mathbf p}_{k}$
respectively, and the environment state is ${\mathbf x}_{j}$, the ${\mathcal T}$ predicts the observed
outcome as ${\mathbf t}_{j}$.\  All quantities typeset in boldface are interval objects, usually just intervals,
but they may be interval vectors as well.\  Contrary to the earlier theoretical attempts (for the relevant
references see the literature cited in \cite{part1}) we no longer insist that
experimental intervals ${\mathbf m}_{j}$ are guaranteed, i.e. that they contain the true values with
probability equal exactly to $1$, nevertheless they may have this property.\  

\medskip
There are essentially two goals addressed by uncertain data processing:

\begin{itemize}
\item  to determine the values of interesting parameters, ${\mathbf p}_{1}, \ldots\, {\mathbf p}_{k}$,
best of all together with their uncertainties, or
\item to test whether a given model of phenomenon under study (theory ${\mathcal T}$) is adequate.
\end{itemize}

We will not go into hypothesis testing but instead will concentrate on finding unknown parameters
given the uncertain experimental information.

\CCLsection{How do we find `best fitted' parameters?}
In \cite{part1} we put forward the idea that the so called `best fits' should be based on the distance
between measured and theoretical values.\  In one dimension, when we compare a~single result
of measurement with the predicted one, and at least one of those quantities is an interval, the
mathematically correct distance is the one valid in the interval space {\mm IR}.\ Starting with
the familiar Moore-Hausdorff distance \cite{Moore}, usually written as
\begin{equation}
d\,({\mathbf a},\,{\mathbf b}) = \max\,\left(\left|\,\underline{a} - \underline{b}\,\right|,\,
\left|\,\overline{a} - \overline{b}\,\right|\,\right),\qquad\ {\mathbf a}, {\mathbf b}\,\in\,\protect{\hbox{\mm IR}},
\quad d\,\in\,\protect{\hbox{\mm R}}
\end{equation}
we finally arrived at the tight interval estimate, ${\boldsymbol\rho}({\mathbf t}, {\mathbf m})$,
of the distance between the theoretical prediction ${\mathbf t}$ and the unknown true result
of a~measurement, hidden somewhere within the interval ${\mathbf m}$:

\begin{itemize}
\item when $\textrm{c}({\mathbf t})\,\in\,{\mathbf m}$:
\begin{equation}
\begin{tabular}{lccl}
lower bound: &  $\underline{\rho}$ &=& $\frac{1}{2}\textrm{w}({\mathbf t})$\\
&&&\\
upper bound: &  $\overline{\rho}$ &=& $\max\,\left[\,d\,({\mathbf t}, \underline{m}),\,
 d\,({\mathbf t}, \overline{m})\,\right]$
\end{tabular}
\label{fin-a}
\end{equation}

\item  when $\textrm{c}({\mathbf t})\,\not\in\,{\mathbf m}$:
\begin{equation}
\begin{tabular}{lccl}
lower bound: &  $\underline{\rho}$ &=&  $\min\,\left[\,d\,({\mathbf t}, \underline{m}),\,
 d\,({\mathbf t}, \overline{m})\,\right]$\\
&&&\\
upper bound: & $\overline{\rho}$ &=& $\max\,\left[\,d\,({\mathbf t}, \underline{m}),\,
 d\,({\mathbf t}, \overline{m})\,\right]$,
\end{tabular}
\label{fin-b}
\end{equation}
\end{itemize}
where $\textrm{c}(\cdot)$ stands for the center of its interval argument,\ 
$\textrm{c}({\mathbf t})=\frac{1}{2}\left(\underline{t}+\overline{t}\right)$, and
$d\,(\cdot, \cdot)$ is a Moore-Hausdorff distance between intervals.\

\medskip
Now, equipped with ${\boldsymbol\rho}$, we can think about the distances in $N$-dimensional
spaces.\  They can be constructed, among other, as the counterparts of the so called $L_{p}$ norms,
generally defined as
\begin{equation}
\|\,{\mathbf x}\,\|_{p} = \left[\, \sum_{j=1}^{N}\,\left| x_{j} \right|^{p}\, \right]^{\frac{1}{p}}.
\label{norms}
\end{equation}
Here every individual $x_{j}$ is a~distance measured along the $j$-th coordinate, and $p$ is a~fixed,
positive real number.\  The most important are norms $L_{1}$, $L_{2}$, and $L_{\infty}$.\ Specifically
we have:

\begin{itemize}
\item {$L_{1}$ distance | a.k.a. Manhattan metric or taxi driver metric:}
\begin{equation}
L_{1}({\mathbf t}, {\mathbf m}) = \sum_{j=1}^{N}\, \frac{{\boldsymbol\rho}({\mathbf t}_{j}, {\mathbf m}_{j})}
{\textrm{w}({\mathbf m}_{j})}
\label{L1}
\end{equation}
This norm is used most often when we suspect the presence of outliers
in experimental data set.\ The corresponding classical procedure bears the name LAD (\underline{L}east
\underline{A}verage/\underline{A}bsolute \underline{D}eviation) optimization.

\item {squared $L_{2}$ norm | squared Euclidean distance:}
\begin{equation}
L_{2}^{2}({\mathbf t}, {\mathbf m}) = \sum_{j=1}^{N} \left[\,\frac{{\boldsymbol\rho}({\mathbf t}_{j},
{\mathbf m}_{j})} {\textrm{w}({\mathbf m}_{j})}\,\right]^{2}
\label{L2}
\end{equation}
We have shown here $L_{2}^{2}$ rather than just $L_{2}$, in order to underline its close relationship with
familiar $\chi^{2}$ functional.\ Minimization of $\chi^{2}$, as it is well known, is an~objective
of the famous LSQ method.\  On the other hand, $L_{2}$ is a~monotonous function of its positive arguments,
so the minima of $L_{2}$ are located at the same arguments as minima of $L_{2}^{2}$.

\item {$L_{\infty}$ or maximum distance | in interval analysis serves as box's diameter:}
\begin{equation}
L_{\infty}({\mathbf t}, {\mathbf m}) = \max_{j=1, \ldots\, N}\, \dfrac{{\boldsymbol\rho}({\mathbf t}_{j},
{\mathbf m}_{j})} {\textrm{w}({\mathbf m}_{j})}
\label{L-8}
\end{equation}
This metric, in turn, is best applicable for calibration purposes \cite{calibration}.\  Here the goal is to
approximate uniformly the set of experimental points via any simple curve (or surface), not necessarily
physically meaningful, but easy to evaluate.
\end{itemize}

In classical data analysis every single functional shown above is treated differently than the remaining ones.\
Contrary, using interval methods, we need not to follow this path and develop procedures specific to each
metric in turn.\  It is entirely possible to use exactly one and the same general purpose procedure to locate the
global minimum of either functional.\  Such a~procedure may be, for example,  similar to that first
described 35 years ago by Skelboe \cite{SSkelboe}, and known as Moore-Skelboe algorithm.

\CCLsection{Troublesome interval output}
Regardless of the interval minimizer we shall use, the final outcome appears almost always troublesome.\
When the result is a~single interval box, then the lengths of its edges are usually much larger
than final uncertainties of the searched parameters as delivered by other methods.\  This is because such
a~result, being an interval hull of what was sought for, contains also many `bad' solutions.\
In fact, the true solutions occupy only a little fraction of the volume returned by algorithm.\
Whatever the reason, our very reliable results simply look poorly, and are by no means competitive.\

\medskip
At the other extreme, when our minimizing algorithm delivers many boxes -- and by many we mean not
two, three or even dozen boxes, but rather hundreds, or maybe even thousands of them -- we are in troubles
again.\
There is no simple way to present such results to other researchers in a~simple, compact form,
acceptable also by publishers.\ Of course, we can quickly calculate the convex hull (or hulls, if the
set of returned boxes is not simply connected) of all boxes, but this takes us back to the previous
situation.

\medskip
Even if our results happen to be quite narrow -- shall we call them `guaranteed?'\  Certainly not, whenever
the input data have a~form of the mean value and standard deviation, as it is most often the case.\

\medskip
Hmmm.\  Let's think again.\  Suppose, we have quite a~number of boxes covering some domain
in parameter space, where the true solutions are located.\  Aren't those points the results of what
is called `indirect measurement?'\  Of course, they are!  If so, then nothing can prevent us from
treating them as usually and calculate their mean values, dispersion, etc.

\CCLsection{Means, variances and correlations}
Suppose the outcome returned by a~minimizing routine is a~cluster of $N_{box}\,$\ simply connected
boxes ${\mathbf B}_{j},$\ $j=1,\,\ldots,\,N_{box}$,\ covering a~single solution.\
How to calculate the ``\/ordinarily\/'' looking answers to our original problem?\  The number of our
indirect measurements is no longer finite, as it takes place in direct measurements.\  It is even
uncountably infinite, but this fact alone is no real obstacle.\  Just in place of various sums we will
have to calculate some definite integrals, that's all.\  During calculations we have to assume that
probability density is uniform in the interiors of all boxes.\  This position may seem strange at first sight
(intervals can never be treated in this spirit!) but is entirely correct.\ The final formulae, valid when\
$N_{box}>1$,\  are following:

\begin{itemize}

\item \textbf{mean values of unknown parameters} | center of gravity of a~cluster
\begin{equation}
{\mathbf p}_{0} =
\frac{\sum_{j=1}^{N_{box}}
\left[\hbox{\textrm\textbf center}\,({\mathbf B}_{j})\times\textrm{Volume}\,({\mathbf B}_{j})\right]}
{\sum_{j=1}^{N_{box}}\, \textrm{Volume}\,({\mathbf B}_{j})},
\label{means}
\end{equation}
where now\ ${\mathbf p}$\ denotes a~real-valued, $k$-dimensional vector of searched parameters, not
their ranges:\ ${\mathbf p} = \left(p_{1}, \ldots,\, p_{k}\right)$,\ and the subscript\ `$0$'\ indicates their
mean (expected) values.\ Of course,\ `$\textrm{center}({\mathbf B}_{j})$'\  is also a~real-valued,
$k$-dimensional vector, pointing -- you guessed  --to the center of box\ ${\mathbf B}_{j}$.\
The meaning of the number\ `$\textrm{Volume}({\mathbf B}_{j})$'\ is self-explanatory.\\

\item \textbf{dispersions (variances) of parameters}\\
We use the textbook definitions for the covariance of two multidimensional random variables
${\mathbf X}$ and ${\mathbf Y}$, when their expected values, ${\mathbf x}_{0}$ and ${\mathbf y}_{0}$,
respectively, are known:
\begin{equation}
\hbox{\textrm Cov}\,({\mathbf X}{\mathbf Y}) = \langle({\mathbf X} - {\mathbf x}_{0})
({\mathbf Y} - {\mathbf y}_{0})\rangle,
\end{equation}
where the braces\ `$\langle\cdot\rangle$'\ mean the average (expected) value.\  The variance of
a~random multidimensional variable can be computed on two equivalent ways, either as
\begin{equation}
\sigma^{2}({\mathbf X}) = \langle( {\mathbf X}-{\mathbf x}_{0})^{2}\rangle\qquad\textrm{or as}\qquad
\sigma^{2}({\mathbf X}) = \textrm{Cov}({\mathbf X}\cdot\,{\mathbf X}).
\end{equation}
In our case ${\mathbf X}={\mathbf Y}=(p_{1}, \ldots, p_{k})$.\  If we denote the range (interval) of
parameter $p_{m}$ as ${\mathbf x}$, and the range of parameter $p_{n}$ as ${\mathbf y}$,
\underline{both limited to current box under study}, as indicated by the summation index $j$,\
then the off-diagonal elements of the covariance matrix\ ($m\ne\,n$)\ are expressed as:
\begin{equation}
\textrm{Cov}(p_{m}p_{n}) = 
\frac{\sum_{j=1}^{N_{box}} \Bigl[
(\overline{\mathbf x}-2{\mathbf x}_{0})\,\overline{\mathbf x} 
- (\underline{\mathbf x}-2{\mathbf x}_{0})\,\underline{\mathbf x}\Bigr]_{j}
\Bigl[(\overline{\mathbf y}-2{\mathbf y}_{0})\,\overline{\mathbf y} 
- (\underline{\mathbf y}-2{\mathbf y}_{0})\,\underline{\mathbf y}\Bigr]_{j}\,RV_{xy}
}{4\,\sum_{j=1}^{N_{box}}\,\textrm{Volume}\,({\mathbf B}_{j})},
\label{off-diag-cov}
\end{equation}
where ${\mathbf x}_{0}$ and ${\mathbf y}_{0}$ are mean values of $p_{m}$ and $p_{n}$, respectively,
as computed earlier from (\ref{means}).\  Newly introduced symbol $RV_{xy}$ means `reduced volume',
that is the volume of $k-2$ dimensional box containing all parameters except\ $p_{m}$\ and\
$p_{n}$:
\begin{equation}
RV_{xy}= \prod_{{\mathbf Z}\ne\,p_{m},\ {\mathbf Z}\ne\,p_{n}}\ 
(\overline{\mathbf z}-\underline{\mathbf z})_{j}
\end{equation}

For diagonal elements of the covariance matrix, when $m=n$, we have instead:
\begin{equation}
\sigma^{2}(p_{m}) =
\frac{\sum_{j=1}^{N_{box}} \Bigl[\overline{\mathbf x}^{2}
+\underline{\mathbf x}\overline{\mathbf x}
-3{\mathbf x}_{0}\overline{\mathbf x}
+\underline{\mathbf x}^{2}
-3{\mathbf x}_{0}\underline{\mathbf x}
+3{\mathbf x}_{0}^{2} \Bigr]_{j}\times\textrm{Volume}({\mathbf B}_{j})}
{3\,\sum_{j=1}^{N_{box}}\textrm{Volume}({\mathbf B}_{j})}
\label{diag-cov}
\end{equation}\\

\item \textbf{correlations between parameters}\\
According to any textbook on statistics, coefficient of correlation between any two multidimensional
random variables is defined as:
\begin{equation}
\rho_{\textrm xy} = \frac{\textrm{Cov}({\mathbf X}{\mathbf Y})}
{\sigma({\mathbf X})\cdot\sigma({\mathbf Y})}
\end{equation}
There should be no problem with calculating this quantity when we already have all necessary
ingredients, obtained from (\ref{means}), (\ref{off-diag-cov}) and (\ref{diag-cov}).
\end{itemize}

\CCLsection{Discussion}
Omission of the case $N_{box}=1$ was deliberate.\  It is both easy and hopeless case.
And here is why.\  Easy part consists in calculating the mean values of unknown parameters.\
They all are simply equal to the centers of corresponding ranges.\  It also easy to show
that their dispersions have to be equal to halves of the widths of their ranges.\  One will be
nevertheless strongly disappointed with correlations between parameters: they are none,
equal exactly to zero.\ But could all this be true?\ Certainly not.

\medskip
The natural question is how accurate are the suggested here results.\  The boxes comprising
the simply connected set covering the domain of possible solutions are not all created equal.\
Some of are them completely filled with the possible solutions, while others, those located at
the boundaries, are filled with solutions only in part.\  This must necessarily affect our results,
since those were derived with only the first kind of boxes in mind.\  It is intuitively clear that
the more boxes we have, and the smaller they are, the `filling factor' will be closer to $100\,$\%.\
Consequently, our results will be closer to reality.\  All we can say is that the dispersions should
come out always overestimated.\  For the cases where both input data and the theory are
correct, that is.\  In statistical language we may say that our estimate of dispersions
(or variances, if you prefer) is consistent but positively biased.\  Fortunately, this makes no
harm.\ 
 
\medskip
Quite a~different story concerns covariances and correlations.\ As we could see, our ignorance
in that matter remains completely intact, when we have at our disposal only a~single box.\
Of course, increasing the number of boxes will take us closer to the true values.\  In case
of off-diagonal elements of covariance/correlation matrix we have no guarantee that convergence
will be one-sided.\  This brings us to the question how many boxes do we really need?\
The exhaustive answer to this problem is beyond current author's capabilities.\  One may hope,
with analogy to other statistical problems, that sensible results should start to appear when
$N_{box}$ exceeds, say $20$.\  Fortunately, the optimizing routine usually delivers much more
boxes, counted in hundreds.

\medskip
We haven't discussed the question of complexity in this paper.\  From what was said, it is clear
that better, more accurate results, are also more costly than just rough estimates:\ depending
on whether we are working with a single box or with many boxes.

\CCLsection{Conclusions}
Interval-oriented routines not only generate reliable estimates of unknown parameters as a~result
of uncertain data processing.\  So obtained results {\em can\/} be safely and reliably `translated'
into more widespread statistical form of presentation.\

\medskip
Interval perspective sheds completely new light on experimental data processing.\  Here we see
with details what is in reality going on.\  Moreover, in many cases interval methods allow for objective
estimates of accuracies, with no need for human experts (who sometimes err very much
in their estimates).

\CCLsection*{Acknowledgments}
This work was done as a~part of author's statutory activities at  the  Institute  of Physics,
Polish Academy of Sciences.


\begin{thebibliography}{99}

\bibitem{part1} Marek W.~Gutowski {\em Breakthrough in Interval Data Fitting.~\ 
I.~The Role of Hausdorff Distance\/}, in this proceedings.

\bibitem{Moore}
R.E.~Moore
{\em Interval Analysis\/}
Prentice Hall, Englewood Cliffs, NJ, 1966

\bibitem{calibration}
A.~Voschinin, N~Skibitski, {\em  Interval calibration model of multisensor system\/},
Intelligent Data Acquisition and Advanced Computing Systems: Technology and Applications,
IDAACS'2003. Proceedings of the Second IEEE International Workshop on, pp.~253--256

\bibitem{SSkelboe}  S.~Skelboe,   {\em   Computation   of   rational
 interval functions\/}, BIT \textbf{14}, 87---95, 1974
 
\end{thebibliography}
\end{document}